\documentclass[fleqn,twoside]{article}
\usepackage{espcrc2}

\usepackage{graphicx}
\usepackage[figuresright]{rotating}

\newcommand{\AmS}{{\protect\the\textfont2
  A\kern-.1667em\lower.5ex\hbox{M}\kern-.125emS}}

\title{Investigating Dark Energy with Black Hole Binaries}

\author{Laura Mersini-Houghton\address[UNCCH]{Department of Physics and Astronomy, The University of North Carolina at Chapel Hill, Phillips Hall, CB \# 3255, Chapel Hill, NC 27599-3255, USA}\address[DAMTP]{DAMTP, Center for Mathematical Sciences, University of Cambridge, Wilberforce Road, Cambridge CB3 0WA, UK}%\address[MCSD]{Mathematics and Computer Science Section, 
        \\Adam Kelleher \addressmark[UNCCH]\addressmark[DAMTP]}
\begin{document}

\begin{abstract}
The accelerated expansion of the universe is ascribed to the existence of dark energy. Black holes accretion of dark energy induces a mass change proportional to the energy density and pressure of the background dark energy fluid. The time scale during which the mass of black holes changes considerably is too long relative to the age of the universe, thus beyond detection possibilities. We propose to take advantage of the modified black hole masses for exploring the equation of state $w[z]$ of dark energy, by investigating the evolution of supermassive black hole binaries on a dark energy background. Deriving the signatures of dark energy accretion on the evolution of binaries, we find that dark energy imprints on the emitted gravitational radiation and on the changes in the orbital radius of the binary can be within detection limits for certain supermassive black hole binaries. In this talk I describe how binaries can provide a useful tool in obtaining complementary information on the nature of dark energy, based on the work done with A.Kelleher.       
\vspace{1pc}
\end{abstract}

%\begin{document}

%\begin{abstract}
%The accelerated expansion of the universe is ascribed to the existence of dark energy. Black holes accrete dark energy. The accretion induces a mass change proportional to the energy density and pressure of the background dark energy fluid. The time scale during which the mass of black holes changes considerably is long relative to the age of the universe, thus beyond detection possibilities. We propose to take advantage of the modified black hole masses for exploring the equation of state $w[z]$ of dark energy, by investigating the evolution of supermassive black hole binaries on a dark energy background. Deriving the signatures of dark energy accretion on the evolution of binaries, we find that dark energy imprints on the emitted gravitational radiation and on the changes in the orbital radius of the binary can be within detection limits for certain supermassive black hole binaries. This talk describes how binaries can provide a useful tool in obtaining complementary information on the nature of dark energy.       
%\vspace{1pc}
%\end{abstract}

\maketitle

\section{INTRODUCTION}

Dark energy drives the universe into an accelerated expansion. A decade after its discovery, the nature and origin of dark energy remain as elusive as ever. The detection of time variations in the equation of state $w[z]$ of dark energy, given by the ratio $w[z] = p /\rho$ of its pressure $p$ to the energy density $\rho$, is notoriously hard to constrain. A precise measurement of $w[z]$ will guide our theoretical exploration and shed light into the type of dark energy, thereby helping us understand our present universe and its future evolution. For these reasons, a precision measurement of $w[z]$ is of the highest priority in cosmology. Our current bounds are derived by a combination of data from cosmic microwave background radiation (CMBR), Large Scale Structure (LSS) surveys, baryon acoustic oscillation (BAO) and Supernova $1A$ (Sn1A) data.  Current data analysis constrains the dark energy equation of state to $-0.14 < 1+w < 0.12$ \cite{WMAP5yr}.  While this represents progress, we would like to know if $1+w$ is positive, negative or $0$, and whether or not w changes with time. Each case corresponds to quite different types of this mysterious energy and, it leads to dramatically different predictions for the future evolution of the universe. 

Dark energy evolves slowly with time for the cases when $1+w$ is positive or negative. However, since the bounds on its equation of state are close to a pure cosmological constant, $w=-1$, then unlike matter dark energy does not cluster. For these reasons a universe filled with dark energy can be reasonably assumed to be a background of a perfect cosmic fluid. T.Jacobson and other authors \cite{jacobson,babichev1} took dark energy to be a cosmic 'fluid' and showed that the mass of a black hole changes due to accretion of this background 'fluid'.  The mass of the black hole increases or decreases, depending on the sign of $1+w$.  This effect would be hard to observe in a single black hole, as the time scale for this phenomena is quite slow relative to the age of the universe. For example the evolution time scale $\tau$ for a solar mass black hole is about $10^{32} yrs$. We proposed in \cite{ourpaper} to use the evolution of supermassive black hole binaries instead of single black hole accretion to probe dark energy. The reason for our proposed method relies on the fact that, instead of measurements of the evolution time scale of a single black hole, the 'footprints' of $w[z]$ for supermassive black hole binaries can be observed and tracked down through the modifications introduced by dark energy accretion in the  the orbital separation of the binary and its emitted gravitational waves. For certain binaries dark energy modifications can be detectable by gravitational waves or X-ray and radio measurements. The information obtained by these modifications should help increase our bounds on $w[z]$.  The use of binaries for probing dark energy is a different approach from our current methods of large scale experiments, since the observation utilizes localized systems, thereby avoiding noise inherited by propagation of signals through the vast structures of the universe.

\section{Dark Energy Accretion by Black Holes}

\subsection{Evolution of a Single Black Hole on the Dark Energy Background: Review}
Let us assume that dark energy corresponds to a perfect fluid with an energy density $\rho$, pressure $p$ and equation of state $w[z]$ as a function of the redshift $z$, related by $p = w[z] \rho$. The case of a pure vacuum energy would have $\rho = -p$. As long as $w[z]$ is close to -1 and varies slowly with time, the metric solution from Einstein's equations is taken to be approximately that of a De Sitter geometry, i.e. for $w=-1$.

A particularly interesting case is that of a Schwarzschild black hole in the background of the dark energy cosmic fluid.  This case was studied in \cite{jacobson,michel,babichev1,babichev2} and the fluid accretion flowing through the black hole horizon was solved analytically. The perfect fluid energy-momentum tensor with equation of state $p = w\rho$ is assumed to be
\begin{equation}
T^{\mu\nu} = \rho(1+w)u^{\mu}u^{\nu} + w\rho g^{\mu\nu}
\end{equation} 
where $g^{\mu\nu}$ is the inverse of the Schwarzschild metric, $g_{\mu\nu} = diag(-(1-2M/r),(1-2M/r)^{-1},r^{2},r^{2}sin^{2}\theta)$, and for the fluid 4-velocity, $u^{\mu}u_{\mu} = -1$. Integrating the energy-momentum conservation equation and the projection of the fluid 4-momentum into it leads to the expression 
\begin{equation}
u \left(\frac{M}{r}\right)^{2} exp\left[\int_{\rho_{\infty}}^{\rho}\frac{d\rho'}{\rho'+p(\rho')} \right]=-A
\end{equation}
and
\begin{eqnarray}
\left(\rho + p \right) \!\left( 1 - \frac{2M}{r} + u^{2} \right) ^{1/2} \!\! \\ \nonumber
\times exp\left[ - \int_{\rho_{\infty}}^{\rho}\frac{d\rho'}{\rho'+p(\rho')} \right] = C
\end{eqnarray}

Following Babichev et al \cite{babichev2}, these expressions are manipulated into one for $r^{2}T_{0}^{r}$.  Then, integrating the conservation law over the volume within the event horizon yields 
\begin{equation}
\dot{M} = -4 \pi r^{2} T_{0}^{r} = 4 \pi A M^{2} \left[ \rho_{\infty} + p(\rho_{\infty})\right]
\label{massrate}
\end{equation}

This expression can be integrated to give M as a function of time, neglecting the cosmological time evolution of $\rho_{\infty}$:
\begin{equation}
M(t) = \frac{M(0)}{1 - \frac{t}{\tau}}.
\label{bhde}
\end{equation}
The timescale for the accretion of dark energy Eq. \ref{bhde} is given by the parameter $\tau$, and is $\tau = \frac{1}{( 4 \pi A M(0)[\rho_{\infty} + p(\rho_{\infty})])}$.  For a black hole of a mass $m = a m_s$ which is $a$ times larger than a solar mass $m_s$, the evolution timescale is roughly $\tau = 10^{32}/a$ years.  This is usually much longer than the age of the universe, thus beyond observational feasibility. But the detection possibilities of the dark energy accretion improve dramatically for the case of black hole binaries. The reason is that the evolution of the black hole binaries in the background of dark energy is different from that of a single hole. Therefore, as described below, detection of the modifications to the orbital radius and the emission of gravitational radiation from these binaries in the background of the dark energy fluid is within reach, \cite{ourpaper} .

\subsection{Evolution of Black Hole Binaries in the Background of Dark Energy}

The Hulse-Taylor effect predicts a decrease of the orbital radius of the binary due the energy lost by the emission of gravitational radiation. This prediction has been succesfully tested. 
The evolution of black hole binaries in the background of dark energy is modified relative to the Hulse-Taylor effect, due to the accretion of dark energy by the stars in the binary. Since the change in the mass of the stars, Eq.\ref{bhde}, has a direct dependence on the parameters of the dark energy being accreted, specifically on $w[z]$, then this information is carried out on the amplitude and the power of gravitational radiation produced \cite{ourpaper} by the binary. The dark energy accretion also imprints modifications in the orbital separation and merging time. 

The modifications induced from the background dark energy fluid onto the evolution of the binary, namely on the frequency $\omega$ of gravitational waves emitted by the binary and, on the orbital separation $R$, can be derived as follows: the change of the gravitational energy of the binary is equal to the power lost due to gravitational radiation\cite{BHpower}:
\begin{equation}
\frac{d}{dt}\left( m_{1} + m_{2} - \frac{1}{2}\frac{m_{1}m_{2}}{R}\right) = P_{GW}
\end{equation}
where 
\begin{equation}
P_{GW} = \frac{-32 G^{4}}{5 c^{5}} \left[ \frac{m_{1}^{2}m_{2}^{2}(m_{1} + m_{2})}{R^{5}}\right]
\end{equation}

This is a temporal equation. Power losses via emission of gravitational radiation drives the binary's configuration to a new gravitational equlibrium separation. As a result the orbital radius $R$ decreases and eventually the stars inspiral and merge.
For binaries immersed in the dark energy fluid, it should be noticed that the point of gravitationally stable configurations of the binary at each moment is now driven by two effects: the usual loss of energy via gravitational waves emission; and, the stars changing mass (leading to a change of the gravitational energy of the binary) due to dark energy accretion. The masses of the two black holes in the binary, $m_1$ and $m_2$, are increasing or decreasing with time, Eq.\ref{bhde}, depending on $(1+w)$ being positive or negative. Therefore the two terms that induce changes in the orbital radius and period, namely: modifications due to dark energy accretion, and modifications due to energy losses from gravitational radiation compete with each other and determine the evolution of the orbit.

Without loss of generality, this expression can be algebraically simplified by taking the two stars to be of equal mass, $m_{1} = m_{2} = m$ with $m_0 = m(0)$.  Replacing the constant mass of a black hole with the new expression, the mass rate of change of the black holes induced by the dark energy accretion from Eq.\ref{bhde}, leads to a differential equation for the evolution of the binary R(t).  
\begin{eqnarray}
{}R^{3} \frac{dR}{dt} = -\frac{64}{5} \frac{2 m_{0}^{3}}{\left[1 - 2Ltm_{0} \right]^{3}} \\ \nonumber{}- \!\!\left[\frac{-4LR^{4}m_{0}}{\left[1 - 2Ltm_{0} \right]} + 8 L R^{6} \right]
\label{diffR}
\end{eqnarray}
where the parameter $L$ denotes $L = \frac{c^{3}}{2 G^{2}m_{0} \tau }$. This parameter contains all the modifications induced by the dark energy background and the new modification terms due to dark energy to the orbital radius can be tracked down from all the terms in Eqs.\ref{solR} that contain $L$. It should be noted that $L \simeq (1+w)$ is positive for quintessence type fluids; $L$ is negative for phantom type fluids ($1+w <0$); and, it becomes identically zero for $w=-1$. In the latter case, all the modifications due to dark energy on the orbital radius vanish. The sign of $L \ne 0$ determines the behaviour of the orbit, i.e. whether it grows or decreases with time for the cases when the '$L-terms$' dominate over the conventional gravitational waves term.

An approximate solution to Eq.\ref{diffR} for the case when dark energy changes adiabatically is \cite{ourpaper}  
\begin{eqnarray} \label{solR}
 R(t,w)=\!  R_{0} [ 1 + 16 L m_{0} \left( \frac{G^{2}}{c^{3}} \right) t \\ \nonumber - 32 L R_{0} \left( \frac{G}{c}\right)t - \frac{64}{5}\left( 4 \frac{G^{3}}{c^{5}}\right)\left[\frac{8 t m_{0}^{3}}{R_{0}^{4}} \right] ] ^{1/4},
\end{eqnarray}
The fourth term in the expression Eq.\ref{solR} corresponds to the conventional Hulse-Taylor term. As a consistency check, the Hulse-Taylor equation is recovered in the limit when our universe approaches a DeSitter geometry $w\rightarrow -1$, (i.e. $L\rightarrow 0$). The Hulse-Taylor term describes the changes in the orbital radius that result for the energy losses of a binary from the emitted gravitational radiation. The terms proportional to $L$ are the new modification terms to the evolution of black hole binaries. They account for the effects of dark energy accretion by the system. It should be noticed that one of the dark energy modification terms is of opposite sign to the Hulse-Taylor term; and, the type of dark energy with $(1+w)$ positive or negative leads to different types of evolution for the binary. An analysis of the solution for $R[z]$ shows that the dark energy terms can dominate the evolution of the orbital radius, Eq.\ref{solR}, for certain cases of supermassive black hole binaries with large separation, quantified below. The different time evolution of the two terms, dark energy and the conventional Hulse-Taylor one, in Eq.\ref{solR} on $R$ and $m$, especially the linear dependence of $\dot{R}$ on the equation of state of dark energy $1+w$, allow us to discriminate among the modifications to the orbit induced by dark energy and for probing the dark energy equation of state $w[z]$ by observing the rate at which the orbit change $\dot{R}$.

In order to quantify the analysis of the above expression Eq.\ref{solR} and discuss the interplay between the dark energy and Hulse-Taylor types of modifications in the orbital radius we can parameterize the binary as follows: let the initial mass of the star, (before modifications due to accretion), be $m_{0} = a m_{s}$ where $m_s$ is a solar mass and $a$ a parameter; and the orbital radius be $\beta$ times larger than the Schwarzchild radius of each star $R_{0} = 2 m_{0} \beta \frac{G}{c^{2}}$.  Then, the ratio of the two correction terms to the binary's orbit $R$ in Eq.\ref{solR}, the Hulse-Taylor correction due to the emission of  gravitational waves (GW), and corrections due to dark energy (DE) accretion (terms containing $L$), is
\begin{equation}
\frac{GW correction}{DEcorrection} = \frac{10^{45}}{(2 \beta)^{5}(1+w)a^{2}} \ge 1
\label{ratio}
\end{equation}
We can use Eq.\ref{ratio} to quantify the classes of binaries for which the dark energy correction terms dominate over the Hulse-Taylor term. Clearly for large enough separations ($\beta$) or masses of the black holes ($a \simeq 10^{18}$) the corrections due to dark energy can dominate over gravitational radiation.

Observing this effect via gravitational waves experiments, we need both parameters of the binary to be such that they favor the dark energy corrections over the Hulse-Taylor corrections to $R$, while at the same time being within observable ranges of frequency windows. The frequency and amplitude of gravitational radiation from these systems are given by 
\begin{equation}
f = \frac{10^5}{(2 \beta)^{3/2} a}
\end{equation}
and
\begin{equation}
h = \frac{1}{r} \frac{2}{\beta} a 10^{3},
\label{frequency}
\end{equation}

where r is the distance of the binary from the observer.  Eq.\ref{frequency} constrains the second parameter $\beta$ not to be too large. For example LIGO is designed to detect radiation around 150 Hz optimally, and with an amplitude greater than $h = 10^{-23} Hz^{1/2}$.\cite{LIGO}  LISA can detect much lower frequency radiation, down to $\sim 10^{-5} Hz$.\cite{LISA}. Since for both GW experiments, the binary radius $\beta$ can not be too large, then our requirement of Eq.\ref{ratio} for using binaries to probe dark energy via their modifications on the orbital radius $R$, can be fulfilled by considering supermassive black holes, $a\gg 1$, (see \cite{ourpaper} for specific examples).

\subsection{Observation Candidates?}

There have now been observations of black hole binary systems, and we can assess the effects of dark energy accretion on these systems.    One example is galaxy 0402+379, observed in 2007 with VLBA.  It has parameters $a = 210^{8}$, $2\beta = 10^{6}$, and $r = 10^{26}m$.  This case, while not observable through gravitational radiation, is one that is sensitive to the sign of $ 1 + w$, and it's merging time is either greatly accelerated (1,000 years instead of 60,000 years) by the effect of dark energy accretion when $1+w > 0$, or it is driven apart when $1 + w < 0$.

Another example is Radio Galaxy OJ287, observed in 2008 with VLBA \cite{VLBA}.  This system is more complicated, since one of the black holes is more massive than the other.  The parameters for this system are $R_{0} \simeq 10^{20}m$, or $2\beta \simeq 10^{7}$ and $r \simeq 3.5 Mly \simeq 10^{22} m$.  While it is straightforward to treat this system with the above equations for $m_{1} \neq m_{2}$, the point is well illustrated by a simpler system of comparable parameters.  We will use $a \simeq 10^{9}$.  This gives $f \simeq 10^{-9} Hz$ and $h \simeq 10^-20$.  Again, this system falls outside of observable ranges for gravitational radiation, but the merging time is shortened by three orders of magnitude when $1 + w > 0$, and the black holes are pulled apart for $1 + w < 0$.
\section{Conclusions}
%\subsection{Results}
It is remarkable that localized systems like black hole binaries can be used to provide information about dark energy. The method described here provides a complementary way of probing the dark energy equation of state by using supermassive black hole binaries. Its strength lies on the fact that it avoids the noise inherited by the signal propagating through the vastness of structure in the universe and it takes advantage of existing experiments, initially designed to investigate gravitational waves or structure. Supermassive black hole binaries could soon help to shed light on the nature of dark energy.  

By observing gravitational radiation from black hole binaries, we might distinguish the waveform from a system accreting dark energy from one that does not. Observing changes in the orbital radius over a fraction of the binary's period with X-ray and Radio measurement is entirely possible with our current experiments and provides a wealth of information on $w[z]$ through the dark energy correction terms in Eq.8,9.
 This method should helps us pinpoint at least whether dark energy is a quintessence or a phantom type, or simply a cosmological constant.  Observing how the waveform differs from the cosmological constant case gives further information about the sign of $1 + w$ \cite{eos}.  If $1 + w > 0$, the masses of the black holes will increase, they will spiral in faster, and this will result also in a faster increase in frequency of their gravitational radiation.  If $1 + w < 0$ \cite{phantom}, the masses decrease and the system can be pulled apart by the effects of dark energy accretion.

It is possible we are close to collecting evidence from existing observed supermassive black hole binaries that the phantom type $(1+w)<0$ which rips the binary apart may be already disfavored. One such binary of supermassive black holes was recently observed \cite{borosin}. An interesting question is whether this method can be used to test theories of modified gravity and to discriminate those from the dark energy models. The evolution of binaries on the background of modified gravity is currently under investigation.

\end{document}